\documentclass[12pt]{article}
\usepackage{graphicx}
\usepackage{epsfig}
\usepackage{amsbsy}

\begin{document}
\makeatletter \Large \baselineskip 10mm

\centerline{\bf Laser-induced operations with charge qubits}
\centerline{\bf  in a double-well nanostructure}

\vskip 5mm

\centerline{A. N. Voronko, L. A. Openov$^{*}$}

\vskip 5mm

\centerline{\it Moscow Engineering Physics Institute} \centerline{\it
(State University),115409 Moscow, Russia}

\vskip 5mm

$^{*}$e-mail: LAOpenov@mephi.ru

\vskip 5mm

\centerline{\bf Abstract}
We present the results of theoretical studies on operations with
charge qubits in the system composed of two tunnel-coupled semiconductor
quantum dots whose two lowest states (localized in different dots) define the logical qubit states while two excited states (delocalized between the dots) serve for the electron transfer from one dot to another under the influence of the laser pulse. It is shown that in the case of small energy separation between the excited levels, the optimal (from the viewpoint of minimal single-qubit operation time and maximum operation fidelity) strategy is to tune the laser frequency between the excited levels. The pulse parameters for implementation of the quantum NOT operation are determined. Analytical results obtained in the rotating-wave approximation are confirmed by rigorous numerical calculations.

\vskip 5mm


\newpage

\centerline{\bf 1. Introduction}

In attempts to construct a scalable quantum computer, solid state nanostructures are considered as the base for the carriers of quantum information \cite{Nielsen, Valiev}. A progress in the technology of quantum dots fabrication \cite{Jacak} and techniques of controlled incorporation of impurity atoms into semiconductors \cite{Schofield} provide a means to develop various nanostructures with predetermined characteristics. Modern experimental tools allow for manipulation of charge carriers in such systems \cite{Hayashi, Petta, Gorman, Hollenberg} and corresponding measurements \cite{DiCarlo}. For example, a Pauli-spin-blockade transport through a silicon double quantum dot was recently demonstrated \cite{Liu}.

Both nuclear (spin) and electron (spin or charge) states were suggested as logical states of quantum bits (qubits) \cite{Kane, Loss, Openov, Pashkin}. In particular, much attention is now paid to the charge qubits based on the tunnel-coupled quantum dots \cite{Hayashi, Petta, Gorman, Openov, Fedichkin} or a singly ionized pair of donor atoms nearby the surface \cite{Hollenberg, Barrett, Fedichkin2, Openov2}, the basis qubit states $|0\rangle$ and $|1\rangle$ being the orbital electron states $|L\rangle$ and $|R\rangle$ localized in different (left and right) quantum dots or at different donor sites.
Operations with such qubits can be implemented, first, through the changes in the energies $E_{L}$, $E_{R}$ of the states  $|L\rangle$, $|R\rangle$ and the height of the barrier separating the minima of the double-well potential (by the adiabatic changes in the voltages at corresponding gates \cite{Hayashi, Petta, Gorman, Hollenberg}) and, second, through the influence of the resonant (tuned to the resonance with one of the excited levels that plays a role of the {}``transport'' level) or nonresonant laser pulse \cite{Openov, Openov2, Oh, Tsukanov, Paspalakis, Openov3, Tsukanov2, Tsukanov3, Kosionis, Basharov}.

Since the decoherence time $\tau$ of charge states in the solid-state systems is very short even at low temperatures (usually $\tau\sim 1$ ns \cite{Hayashi, Petta}), apart from a specific type of operations with the charge qubits, there is a limitation $t_{op}\ll\tau$ on the time $t_{op}$ of single-qubit operations \cite{Barrett, Fedichkin2, Openov2, Openov4, Stavrou, Cao, Openov5}. The main advantage of operations implemented via  the resonant laser pulses is the small value of $t_{op}$ \cite{Openov, Openov2}. However, in this case there appear additional (and rather severe) requirements to the electron energy spectrum and pulse parameters \cite{Openov, Tsukanov, Tsukanov2, Basharov} which are necessary to increase the operation fidelity. One such requirement is the absence of other energy levels in the vicinity of the transport level since the population leakage to those levels is possible. However, contrary to adiabatic transitions controlled by the gate potentials, the laser-induced operations with qubits are based on the use of relatively distant from each other and/or separated by the high potential barrier centers of electron localization (quantum dots or impurity atoms), so that  the direct tunneling between the ground states of those centers is strongly suppressed \cite{Openov}. This, in turn, results in  bunching of excited levels lying not very close to the continuum into multiplets. The number of levels in a given multiplet depends on the degeneracy of excited states of isolated quantum dots from which this multiplet is formed upon their hybridization. The energy separation between the levels in a multiplet can be very small, resulting in a drastic decrease of the operation fidelity. The energy spectrum of a singly ionized pair of donor atoms in silicon is even more complex because of both the strong degeneracy of excited states of each donor considered as a hydrogen-like atom and the six-fold degeneracy of the silicon conduction band. In the case of nonresonant pulses, the proximity of excited levels to each other is not detrimental, but the value of $t_{op}$ greatly increases \cite{Paspalakis, Tsukanov2, Tsukanov3, Basharov}.

In this work, in order to determine the optimal (from the viewpoint of maximal operation fidelity and minimal operation time) values of the pulse parameters (frequency, duration, and intensity), we consider a model system where, except the ground states $|0\rangle = |L\rangle$ and $|1\rangle = |R\rangle$ (the basis states of the qubit), there is only one pair of excited states, $|2\rangle$ and $|3\rangle$, formed upon hybridization of the corresponding excited states of quantum dots and close to each other in energy. As will be shown below, the problem of laser-induced transitions $|0\rangle \rightarrow |1\rangle$, $|1\rangle \rightarrow |0\rangle$ in such a four-level system allows for an exact solution within the standard rotating wave approximation (RWA), without additional approximations like the adiabatic elimination of high-lying levels \cite{Allen}. Relatively simple analytical expressions for the dependencies of the state amplitudes on time not only help to analyze the results of numerical calculations on much more complex systems \cite{Tsukanov3, Voronko} but enable to find out a novel regime of laser-iduced operations with charge qubits that differs from both resonant and nonresonant (in a commonly accepted interpretation \cite{Tsukanov2, Tsukanov3, Basharov}) regimes.

The paper is organized as follows. In Section 2, the model is described and general analytical solution is obtained using the RWA. In Sections 3-5, particular cases of resonant, nonresonant, and quasiresonant operations with charge qubits are considered, operation times are found, and the results of numerical calculations beyond RWA are presented. In Section 6, concluding remarks are given.

\vskip 10mm

\centerline{\bf 2. Description of the model and general solution}

The energy spectrum of a four-level nanostructure with the symmetric double-well potential is shown schematically in Fig. 1. The one-electron states  $|k\rangle$ are the eigenstates of the Hamiltonian $\hat{H_0} = -(1/2m^\ast)\Delta + U(\mathbf{r})$ where $U(\mathbf{r})$ is the potential energy of an electron in the nanostructure (see Fig. 1), $m^\ast$ is the electron effective mass (hereafter the Planck constant $\hbar=1$):
\begin{equation}\label{eq1}
    \hat{H_0} |k\rangle = \varepsilon_k |k\rangle, \quad k = 0\div3.
\end{equation}
Two lowest states, $|0\rangle$ and  $|1\rangle$, are degenerate in energy. Their wave functions are localized in the left and right wells, respectively, and overlap very weakly. The energies $\varepsilon_2$ and $\varepsilon_3$ of the excited states are closely spaced, and their wave functions are delocalized over the nanostructure. At $t=0$, an electron is in the superposition of $|0\rangle$ and $|1\rangle$ states:
\begin{equation}\label{eq2}
    |\Psi(0)\rangle = \alpha |0\rangle + \beta |1\rangle,
\end{equation}
where $|\alpha|^2 + |\beta|^2 = 1$.

In the presence of the ac electric field $\mathbf{E}(t) =
\mathbf{E}_0\cos{\omega t}$, the Hamiltonian of the system in the dipole approximation has the form:
\begin{equation}\label{eq3}
    \hat{H}(t) = \hat{H}_0 - e\mathbf{E}(t)\mathbf{r},
\end{equation}
where $e$ is the electron charge. We seek the solution of the nonstationary Schr\"{o}dinger equation
\begin{equation}\label{eq4}
    i\frac{\partial |\Psi(t)\rangle}{\partial t} = \hat{H}(t) |\Psi(t)\rangle
\end{equation}
for the state vector $| \Psi(t) \rangle$ in the form
\begin{equation}\label{eq5}
    |\Psi(t)\rangle = \sum^{3}_{k=0}{C_k(t) e^{-i \varepsilon_k t} |k\rangle}.
\end{equation}
Substituting Eq. (\ref{eq5}) into Eq. (\ref{eq4}) and taking Eqs. (\ref{eq1})--(\ref{eq3}) into account, we arrive at the system of coupled differential equations for coefficients $C_k(t)$:
\begin{equation}\label{eq6}
    i\frac{d C_k(t)}{d t} = \frac{1}{2}\sum^{3}_{m=0}{C_m(t) \lambda_{km} e^{i (\varepsilon_k - \varepsilon_m) t}
    (e^{i \omega t} + e^{- i \omega t})}, \quad k = 0 \div 3,
\end{equation}
with the initial conditions
\begin{equation}\label{eq7}
    C_0(0) = \alpha, C_1(0) = \beta, C_2(0) = C_3(0) = 0.
\end{equation}
The values of $\lambda_{km}$ in Eq. (\ref{eq6}) are equal to
\begin{equation}\label{eq8}
    \lambda_{km} = \mathbf{d}_{km} \mathbf{E}_0 = \langle k | -e
    \mathbf{r} | m \rangle \mathbf{E_0},
\end{equation}
where $\mathbf{d}_{km}$ are matrix elements of the operator of dipole moment.

Making use of the rotating wave approximaion (RWA) which is valid at values of
$\lambda_{km}$ and detunings $\delta_2 = \varepsilon_0 + \omega -
\varepsilon_2$, $\delta_3 = \varepsilon_0 + \omega - \varepsilon_3$ (see Fig.1)
much smaller than the frequency of the electric field $\omega$, and introducing the notations $\tilde{C}_2(t) = C_2(t) e^{i \delta_2 t}$, $\tilde{C}_3(t) = C_3(t) e^{i \delta_3 t}$, we have:
\begin{equation}\label{eq9}
    \left\lbrace
    \begin{array}{l}
        i\frac{d C_0 \left( t \right)}{d t} = \frac{\lambda_{02}}{2}\tilde{C}_2 \left( t \right) + \frac{\lambda_{03}}{2}\tilde{C}_3 \left( t \right), \\
        i\frac{d C_1 \left( t \right)}{d t} = \frac{\lambda_{12}}{2}\tilde{C}_2 \left( t \right) + \frac{\lambda_{13}}{2}\tilde{C}_3 \left( t \right), \\
        i\frac{d \tilde{C}_2 \left( t \right)}{d t} = \frac{\lambda_{20}}{2} C_0 \left( t \right) + \frac{\lambda_{21}}{2} C_1 \left( t \right) - \delta_2 \tilde{C}_2 \left( t \right), \\
        i\frac{d \tilde{C}_3 \left( t \right)}{d t} = \frac{\lambda_{30}}{2} C_0 \left( t \right) + \frac{\lambda_{31}}{2} C_1 \left( t \right) - \delta_3 \tilde{C}_3 \left( t \right). \\
    \end{array}
    \right.
\end{equation}
For definiteness, let us view the states $|2\rangle$ and $|3\rangle$ as, respectively, the symmetric and antisymmetric combination of excited states $|\tilde{L}\rangle$ and $|\tilde{R}\rangle$ of isolated quantum dots, $|2\rangle = \left(|\tilde{L}\rangle + |\tilde{R}\rangle \right) / \sqrt{2}$ and $|3\rangle = \left( |\tilde{L}\rangle - |\tilde{R}\rangle \right)
/ \sqrt{2}$. The wave functions of the states $|\tilde{L}\rangle$ and $|\tilde{R}\rangle$ (like those of the states $|L\rangle$ and $|R\rangle$) have similar functional form but are centered in different minima of the double-well potential. Neglecting the overlap integrals that are exponentially small in the parameters $R_0 / a$ and $R_0 / \tilde{a}$ where $R_0$ is the distance between the potential minima, $a$ and $\tilde{a}$ are the localization lengths of the wave functions of the states $|L\rangle$ and $|\tilde{L}\rangle$ respectively, we have $\lambda_{02}=\lambda_{03}=\lambda_{12}=-\lambda_{13}\equiv\lambda
= \langle L|-e \mathbf{r} |\tilde{L}\rangle \mathbf{E}_0 /
\sqrt{2}$ (we suppose that the symmetries of the states $|L\rangle$ and $|\tilde{L}\rangle$  and the electric field polarization are such that $\lambda \neq 0$). Solving the system (\ref{eq9}) with the initial conditions (\ref{eq7}), we finally have the following expressions for the coefficients $C_k(t)$:
\begin{equation}\label{eq10}
    \left\lbrace
    \begin{array}{l}
        C_0(t) = \frac{1}{2} (\alpha + \beta) F_2(t) + \frac{1}{2} (\alpha - \beta) F_3(t), \\
        C_1(t) = \frac{1}{2} (\alpha + \beta) F_2(t) - \frac{1}{2} (\alpha - \beta) F_3(t), \\
        C_2(t) = -i \frac{\lambda}{4 \Omega_2} (\alpha + \beta) \sin(2 \Omega_2 t) e^{-i \frac{\delta_2 t}{2}}, \\
        C_3(t) = -i \frac{\lambda}{4 \Omega_3} (\alpha - \beta) \sin(2 \Omega_3 t) e^{-i \frac{\delta_3 t}{2}}, \\
    \end{array}
    \right.
\end{equation}
where
\begin{equation}\label{eq11}
    F_{2,3} = \left[ \cos(2 \Omega_{2,3} t) - i \frac{\delta_{2,3}}{4 \Omega_{2,3}} \sin(2 \Omega_{2,3} t)
    \right] e^{i \frac{\delta_{2,3} t}{2}},
\end{equation}
\begin{equation}\label{eq12}
    \Omega_{2,3} = \frac{\sqrt{2 \lambda^2 + \delta^2_{2,3}}}{4}.
\end{equation}
In a particular case $\alpha=1$ and $\beta=0$, expressions (\ref{eq10}) -- (\ref{eq12})
coincide with those obtained in Ref. \cite {Tsukanov2}. In what follows, we analyze three different ways of implementation of laser-induced single-qubit operations -- resonant, nonresonant, and quasiresonant.

\vskip 10mm

\centerline{\bf 3. Resonant pulses}

Let the frequency of the laser pulse be tuned to the resonance with the level 2, so that $\delta_2=0$ and $\delta_3=-V$, where $V=\varepsilon_3 - \varepsilon_2$ is the energy separation between the excited states, see Fig. 1. If $V\rightarrow\infty$, this problem is reduced to the issue of laser-induced operations with the charge qubit in the three-level system studied at length in Refs. \cite{Openov,Openov2,Openov3}. Here the level 2 plays the role of the transport level, in that it facilitates the electron transfer $|L\rangle \rightarrow |R\rangle$, $|R\rangle \rightarrow |L\rangle$ between the ground states (i. e., between the logical states of the qubit, $|0\rangle \rightarrow |1\rangle$, $|1\rangle \rightarrow |0\rangle$) but remains unoccupied after the pulse is off if the operation time $t_{op}$ is chosen properly. In this case $\Omega_2 = \lambda \sqrt{2} / 4$, $\Omega_3 \rightarrow \infty$, $F_2(t) = \cos(2 \Omega_2 t)$, $F_3 = 1$, and one has from Eq. (\ref{eq10}):
\begin{equation}\label{eq13}
    \left\lbrace
    \begin{array}{l}
        C_0 \left( t \right) =  \alpha \cos^2(\Omega_2 t) - \beta \sin^2(\Omega_2 t), \\
        C_1 \left( t \right) =  \beta \cos^2(\Omega_2 t) - \alpha \sin^2(\Omega_2 t), \\
        C_2 \left( t \right) = -i \frac{\alpha+\beta}{\sqrt{2}} \sin(2 \Omega_2 t), \\
        C_3 \left( t \right) = 0. \\
    \end{array}
    \right.
\end{equation}
For example, the operation NOT is realized in time $t_{op}^{res} = \pi / 2 \Omega_2 = \pi \sqrt{2}$, i. e., $| \Psi(0) \rangle = \alpha |0 \rangle + \beta | 1 \rangle \rightarrow | \Psi(t_{op}^{res})\rangle = -(\beta | 0 \rangle + \alpha | 1 \rangle)\exp(-i \varepsilon_0 t_{op}^{res})$.

As noted in Ref. \cite{Openov}, even at the exact resonance  $(\delta_2=0)$ the maximum probability $P_{max}$ of the laser-iduced transition
 $|L\rangle \rightarrow |R\rangle$, $|R\rangle \rightarrow |L\rangle$(i. e., the operation fidelity) is close to unity only in the case that other excited levels are well separated from the transport one.
Making use of Eq. (\ref{eq10}) for the probability $P_1(t)$ to find an electron in the state $| 1 \rangle$ at moment $t$ if at $t=0$ it was in the state $| 0 \rangle$, at $V\gg\lambda$
we have
\begin{equation}\label{eq14}
    P_1\left( t \right) = \left|
    C_1\left(t,\alpha=1,\beta=0\right)\right|^2 \approx \sin^4(\Omega_2
    t) - 8 \frac{\Omega_2^2}{V^2} \sin^2(\Omega_2 t)
    \sin^2\left(\frac{Vt}{2}\right),
\end{equation}
and hence
\begin{equation}\label{eq15}
    P_{max} = P_1\left(t_{op}^{res}\right) = 1 -
    \frac{\lambda^2}{V^2} \sin^2{\left(\frac{\pi\sqrt{2}}{2} \cdot
    \frac{V}{\lambda}\right)},
\end{equation}
in agreement with the results obtained in Ref. \cite{Tsukanov2} based on the Hamiltonian
$\hat{H}_0$ written in the basis of the ground and excited states of two isolated quantum dots (and not in basis of stationary states of the double-well nanostructure as in the present work).

In Fig. 2 the function $P_1(t)$ at $V=0.01\omega$ and $\lambda=0.001\omega$ is compared with the numerical solution of the system (\ref{eq3}) beyond RWA.
When choosing the values of the parameters $\lambda_{km}$ for numerical calculations, we took into account the following considerations. If we take the
origin of the coordinates in between the potential minima, then at $R_0 \gg a,
\tilde{a}$ one has from Eq. (\ref{eq8}): $\lambda_{00} \approx
e\mathbf{R}_0\mathbf{E}_0/2$, $\lambda_{01} \approx 0$,
$\lambda_{11} \approx -e\mathbf{R}_0\mathbf{E}_0/2$, $\lambda_{22}
\approx 0$, $\lambda_{23} \approx e\mathbf{R}_0\mathbf{E}_0/2$,
$\lambda_{33} \approx 0$. Since $\lambda_{02} \equiv \lambda
\sim e a E_0$, at $R_0 \gg a$ we have $|\lambda_{00}|,
|\lambda_{11}|, |\lambda_{23}| \gg |\lambda|$ (note that at $a\approx 5$ nm and $\omega = 10$ meV, the value of $\lambda=0.001\omega$ corresponds to the field strength $E_0 \approx 20$ V/cm). From Fig. 2 one can see that in the vicinity of the first maximum of $P_1(t)$, agreement between analytical and numerical results is very good (it is impaired as $t$ increases). The value of $P_{max}$ is 0.9995 and 0.9871 in the former and in the latter case respectively. As $\lambda/V$ and $\lambda/\omega$ decrease, the
deviation of the expression (\ref{eq14}) from the numerical results becomes smaller since
the criteria of RWA and approximations made to derive that expression are fulfilled better. Fig. 3 shows $P_{max}$ versus $V/\lambda$ at $\lambda=0.001\omega$, see Eq. (\ref{eq15}), along with the results of numerical calculations. As expected, the greater is the value of $V/\lambda$, the better is Eq. (\ref{eq15}) applied.

Since in the resonant regime the value of  $t_{op}^{res}\sim 1/\lambda$ increases as $\lambda$ decreases, the condition $\lambda \ll V$ for the high fidelity of operations with the qubits results in the condition $t_{op}^{res} \gg 1/V$, i. e., if the pulse is tuned to the resonance with one of the levels in the multiplet consisting of closely lying levels, the operation time can be very long. For example, at $V \approx \left( 0.01 \div 0.1 \right)$ meV \cite{Voronko} the value of $t_{op}^{res} \gg 0.1$ ns is about the characteristic decoherence time $\tau \sim 1$ ns of charge states in the double-dot nanostructure \cite{Hayashi, Petta}, thus resulting in a drastic decrease of operation fidelity or even impossibility of its implementation.

\vskip 10mm

\centerline{\bf 4. Nonresonant pulses}

Nonresonant electron transitions between the ground states $|0\rangle$ and $|1\rangle$ are realized at large detunings of the laser frequency from the levels 2 and 3, so that
$\lambda \ll |\delta_2|, |\delta_3|$ \cite{Paspalakis, Tsukanov2, Basharov}, see Fig. 1.
If $|\delta_2|, |\delta_3| \ll \omega$, the RWA requirements are met, and from Eqs. (\ref{eq11}), (\ref{eq12}) we have $F_{2,3} = \exp{\left( -i \lambda^2 t / 2
\delta_{2,3}\right)}$. Then from Eq. (\ref{eq10}) it follows that
\begin{equation}\label{eq16}
    \left\lbrace
    \begin{array}{l}
      C_0 \left(t\right) = e^{-i \Lambda_1 t} \left[ \alpha\cos(\Lambda_2 t) + i\beta\sin(\Lambda_2 t) \right] + O\left(|\lambda/\delta_2|\right), \\
      C_1 \left(t\right) = e^{-i \Lambda_1 t} \left[ \beta\cos(\Lambda_2 t) + i\alpha\sin(\Lambda_2 t) \right] + O\left(|\lambda/\delta_3|\right), \\
      C_2 \left(t\right) \approx - \frac{\lambda}{2\delta_2} \left( \alpha + \beta \right) \left( 1-e^{-i \delta_2 t} \right), \\
      C_3 \left(t\right) \approx - \frac{\lambda}{2\delta_3} \left( \alpha - \beta \right) \left( 1-e^{-i \delta_3 t} \right), \\
    \end{array}
    \right.
\end{equation}
where
\begin{equation}\label{eq17}
    \Lambda_1 =
    \frac{\lambda^2\left(\delta_2+\delta_3\right)}{4\delta_2\delta_3},~
    \Lambda_2 = \frac{\lambda^2 V}{4\delta_2\delta_3}
\end{equation}
(here we took into account that $\delta_2-\delta_3=V$). In a specific case
$\alpha=1$, $\beta=0$ our results are consistent with those obtained in Ref. \cite{Tsukanov2}. In particular, it follows from Eq. (\ref{eq16}) that operation NOT,
$|\Psi(0)\rangle = \alpha |0\rangle + \beta |1\rangle \rightarrow |\Psi(t_{op}^{off})\rangle
=$ $i\left(\beta|0\rangle + \alpha|1\rangle\right)$ $\exp(-i
\varepsilon_0 t_{op}^{off} - i \Lambda_1 t_{op}^{off})$ is implemented at
 $t_{op}^{off}=\pi/2|\Lambda_2|=2\pi|\delta_2\delta_3| /\lambda^2 V$ .

As mentioned in Ref. \cite{Tsukanov2}, the optimal strategy (from the viewpoint of minimal $t_{op}^{off}$) is to tune the pulse frequency exactly between the levels 2 and 3, so that $\delta_2 = -\delta_3 = V/2$. But even in this case, the value of $t_{op}^{off}$ is very long, $t_{op}^{off} = \pi V/2 \lambda^2 \gg 1/V$, this time because of the
{}``out-of-resonance condition'' $|\delta_2|, |\delta_3| \gg \lambda$ which
reduces to the condition $\lambda \ll V$ for such a tuning. Moreover, for the same values of $\lambda$ and $V$ the ratio $t_{op}^{off}/t_{op}^{res} = \sqrt{2}V / 4\lambda \gg 1$, i. e., the time of nonresonant transitions is much longer than that of resonant ones. One of the arguments in favour of nonresonant pulses for operations with charge qubits \cite{Tsukanov2} is (along with a smooth temporal evolution of coefficients
$C_0(t)$ and $C_1(t)$ in the state vector $|\Psi(t)\rangle$, see Eq. (\ref{eq5}))
that, as long as the ratios $|\lambda/\delta_2|$ and $|\lambda/\delta_3|$ are small, see Eq. (\ref{eq16}), the excited states remain almost unoccupied during the action of the laser pulse, and hence the decoherence due to relaxation processes $|2\rangle \rightarrow
|0\rangle, |1\rangle$ and $|3\rangle \rightarrow |0\rangle,|1\rangle$ (e. g., upon phonon emission) is strongly suppressed. Note, however, that the decoherence can also stem from the dephasing processes \cite{Fedichkin2}, and hence the large value of $t_{op}^{off}$ is, in our opinion, the drawback of nonresonant operations with charge qubits.

From Fig. 4 one can see that the probability of the $|0\rangle \rightarrow |1\rangle$ transition, $P_1(t) = \left| C_1 \left(t,\alpha=1,\beta=0\right) \right|^2 \approx
\sin^2(\Lambda_2 t)$, obtained from the approximate solution
(\ref{eq16}) at $\delta_2 = - \delta_3 = V/2$ (so that $\Lambda_2
= - \lambda^2/V$), $V=0.01\omega$, and $\lambda=0.001\omega$ agrees well with numerical calculations beyond RWA. This agreement becomes even better as the ratio $\lambda/V$ decreases. Fig. 5 shows the results of numerical calculations of the maximal transition probability $P_{max}$ versus $V/\lambda$ at
$\delta_2 = - \delta_3 = V/2$ and $V=0.01\omega$. The greater the ratio of $V/\lambda$, the closer to unity is $P_{max}$.

\vskip 10mm

\centerline{\bf 5. Quasiresonant pulses}

Above we have considered a specific case of a nonresonant pulse,
$\delta_2 = - \delta_3 = V/2$ (so that the condition $\lambda \ll
|\delta_2|, |\delta_3|$ for applicability of approximations made to derive the expressions (\ref{eq16}) for $C_k(t)$ is $\lambda \ll
V$) and obtained $P_{max} = P_1\left(t_{op}^{off}=\pi V/2\lambda^2
\right) \approx 1$ at $\lambda \ll V$. However from the results of numerical calculations of the maximum transition probability $P_{max}$, see Fig. 5, one can see that there are frequent oscillations of $P_{max}$ as a function of $V/\lambda$ and that the value of $P_{max}$ can be close to unity not only at $\lambda \ll V$ but also at $\lambda\sim V$, i. e., when the {}``out-of-resonance condition'' is not fulfilled. This points to the existence of one more regime of the $|0\rangle \rightarrow |1\rangle$ transition that is different from both resonant and nonresonant regimes (but is analogous to the latter at $\lambda\ll V$). We called it a {}``quasiresonant regime''. In such a regime, just as in the nonresonant regime considered above, the laser frequency is tuned exactly between the levels 2 and 3 (i. e., $\delta_2 = - \delta_3 = V/2$) but, contrary to the nonresonant regime, the condition $\lambda \ll |\delta_2|, |\delta_3|$ (i. e., $\lambda \ll V$) is not necessary.

Restricting ourselves to the case $\alpha=1$, $\beta=0$, at $\delta_2 = - \delta_3 = V/2$ from Eqs. (\ref{eq10}) -- (\ref{eq12}) we have:
\begin{equation}\label{eq18}
    C_1 \left( t,\alpha=1,\beta=0 \right) = - \frac{i}{2\Omega}
        \left[ \tilde{\Omega}_1\sin(\tilde{\Omega}_2 t) -
               \tilde{\Omega}_2\sin(\tilde{\Omega}_1 t) \right],
\end{equation}
where
\begin{equation}\label{eq19}
    \tilde{\Omega}_1 = \Omega + V/4,~
    \tilde{\Omega}_2 = \Omega - V/4,~
    \Omega = \frac{\sqrt{8\lambda^2+V^2}}{4}.
\end{equation}
It follows from Eqs. (\ref{eq18}) and (\ref{eq19}) that the maximum probability
$P_{max} = P_1(t_{op}^{qr}) =\left| C_1\left( t_{op}^{qr},\alpha=1,\beta=0
\right) \right|^2$ of the $|0\rangle \rightarrow |1\rangle$ transition equals to unity if both the conditions $\sin(\tilde{\Omega}_1 t_{op}^{qr}) =
\pm 1$ and $\sin(\tilde{\Omega}_2 t_{op}^{qr}) = \mp 1$ are satisfied, where $t_{op}^{qr}$ is the operation time for the quasiresonant pulse. These conditions result in the following values of $\lambda$ and $t_{op}^{qr}$:
\begin{equation}\label{eq20}
    \lambda = \frac{V \sqrt{2}}{4} \cdot
\frac{\sqrt{\left(4n_1+1\right)\left(4n_2-1\right)}}{\left|2n_1-2n_2+1\right|},
\end{equation}
\begin{equation}\label{eq21}
    t_{op}^{qr} = \frac{2\pi}{V} \left| 2n_1-2n_2+1 \right|,
\end{equation}
where $n_1\geq0$ and $n_2\geq1$ are integers. It follows from Eq. (\ref{eq20}) that $P_{max}=1$ for a discrete set of $\lambda$ values.
The limiting case of the nonresonant regime considered above ($\lambda << V$) corresponds to the values of $n_1=0$ and $n_2 >> 1$. In this case, from Eqs. (\ref{eq20}) and (\ref{eq21}) we have $\lambda\approx V\sqrt{2}/4\sqrt{n_2}$ and $t_{op}^{qr}\approx 4\pi n_2/V$,
so that $t_{op}^{qr}\approx \pi V/2\lambda^2=t_{op}^{off}$.

The expression (\ref{eq20}) for $\lambda$ enables us, on the one hand, to understand the reason of $P_{max}$ oscillations with $V/\lambda$ found numerically in the nonresonant regime ($\lambda \ll V$), see Fig. 5, and, on the other hand, to show a way of how to decrease the operation time. As follows from Eq. (\ref{eq21}), the value of $t_{op}^{qr}$ is minimal (and equal to $2\pi /V$) at $n_1 = n_2$, i. e., at $\lambda = V \sqrt{2(16n^2-1)} / 4$, where $n=1,2,3,...$, or at $n_1=0$ and $n_2=1$ (in this case $\lambda = V \sqrt{6} / 4$). In Fig. 6 the transition probability $P_1(t)$ derived from Eqs. (\ref{eq18}) and (\ref{eq19}) at $\lambda = V \sqrt{6} / 4$ is compared with numerical calculations beyond RWA. One can see that both curves are almost indistinguishable, the value of $P_{max}$ being equal to 1 and 0.999973 in the former and in the latter case respectively.

Let us discuss briefly the situation when the states $|2\rangle$ and $|3\rangle$ are formed upon hybridization of {\it different} excited states of isolated quantum dots ($|\tilde{L}\rangle$ with $|\tilde{R}\rangle$ and
$|\tilde{\tilde{L}}\rangle$ with $|\tilde{\tilde{R}}\rangle$). For definiteness let $|2\rangle = \left(|\tilde{L}\rangle + |\tilde{R}\rangle \right) / \sqrt{2}$ and $|3\rangle = \left( |\tilde{\tilde{L}}\rangle - |\tilde{\tilde{R}}\rangle\right) / \sqrt{2}$. Then $\lambda_{02}=\lambda_{12}=\langle L|-e{\bf r}|\tilde{L}\rangle{\bf E}_0\sqrt{2}\equiv\lambda$, $\lambda_{03}=-\lambda_{13}=\langle L|-e{\bf r}|\tilde{\tilde{L}}\rangle{\bf E}_0\sqrt{2}\equiv\lambda^{\prime} \neq\lambda$. Eqs. (\ref{eq10}) -- (\ref{eq12}) remain valid upon substitution of $\lambda$ by $\lambda^{\prime}$ in expressions for $C_3(t)$ and
$\Omega_3$. Analysis shows that the values of detunings $\delta_2$ and $\delta_3=\delta_2-V$ for which $P_{max}=1$ depend on the value of $|\lambda^{\prime}/\lambda|$, namely: $\delta_2/V=Z+sgn(|\lambda^{\prime}|-|\lambda|)\sqrt{Z^2-Z+1}$ where
$Z=\lambda^2/(\lambda^2-\lambda^{\prime 2})$. So, the energy $\varepsilon_0+\omega$ of the {}``quasiresonant level'' should be closer to the excited state (2 or 3) for which the absolute value of the matrix element of the dipole transition ($|\lambda|$ or $|\lambda^{\prime}|$) is greater. Note that at $\lambda^{\prime}\neq \lambda$ the value of minimum operation time $t_{op}^{qr}=2\pi/V$ appears to be the same as at $\lambda^{\prime}=\lambda$ (and hence, $\delta_2=V/2$). Numerical calculations confirm these results.

\vskip 10mm

\centerline{\bf 6. Conclusions}

Let us summarize the results obtained. If a) the minima of the symmetric double-well potential are separated by a rather high potential barrier so that the ground state of an electron is doubly degenerate (the wave functions of the states $|0\rangle$ and $|1\rangle$
are localized near different potential minima); b) in the upper part of the energy spectrum there is a pair of states, $|2\rangle$ and $|3\rangle$, which lie close in energy and are, respectively, the symmetric and antisymmetric superpositions of excited states of the
left and right well; c) the matrix elements $\langle 0,1|-e\mathbf{r}|2,3
\rangle\mathbf{E}_0=\pm\lambda$ of the dipole transitions are nonzero, then there are three different regimes of the laser-induced operations with a charge qubit (i. e., three regimes of electron transfer between the states $|0\rangle$ and $|1\rangle$ with assistance of {}``auxiliary'' excited states). In the resonant regime, when the laser frequency is tuned to the exact resonance with one of the excited states (e. g., $\varepsilon_0+\omega=\varepsilon_2$), the operation time $t_{op}^{res}=\pi\sqrt{2}/\lambda$ is limited from below by the condition $\lambda \ll V$ where $V = \varepsilon_3 - \varepsilon_2$
is the level separation in a doublet (the operation fidelity is close to unity only if this condition is satisfied). As a result, $t_{op}^{res} \gg 1/V$. In the nonresonant regime, the operation time $t_{op}^{off}$ is minimal if the laser frequency is tuned right between the levels 2 and 3 ($\varepsilon_0 + \omega = \varepsilon_2 + V/2$). In this case $t_{op}^{off} = \pi V/2\lambda^2 \gg 1/V$ because of the {}``out-of-resonance condition'' $\lambda \ll V$ (moreover, $t_{op}^{off} \gg t_{op}^{res}$). In the quasiresonant regime, the laser frequency is also tuned exactly between the states 2 and 3 but now the requirement $\lambda \ll V$ is absent, and hence the minimal operation time $t_{op}^{qr}=2\pi/V$ is about an order of magnitude shorter than in the resonant regime and may appear to be much shorter than the decoherence time for the charge states, $\tau\sim1$ ns (e. g., $t_{op}^{qr}\approx0.04$ ns at $V=0.1$ meV). We stress that all analytical results obtained approximately in different limiting cases are confirmed by numerical calculations beyond RWA.

Finally we note that for the large-scale quantum computations it is necessary that the time $t_{op}$ of one operation was at least four orders of magnitude shorter than the decoherence time $\tau$ \cite{DiVincenzo}. This requirement can be satisfied through decrease in $t_{op}$ and increase in $\tau$. If operations with charge qubits are implemented by means of laser pulses, then in order to lower the value of $t_{op}$ in the resonant regime, the transport level should be chosen so that it was well separated from other excited levels for which the matrix elements of dipole transitions to the ground (localized) states are nonzero, while in the quasiresonant regime one should tune the laser frequency exactly between the levels which are separated from each other as far as possible. As for the decoherence effects, the value of $\tau$ increases upon isolation of the qubit from the leads and the use of capacitance techniques for initialization and measurement \cite{Gorman}.

\newpage

\centerline{\bf Figure captions}

\vskip 4mm

Fig. 1. Schematic view of the potential energy $U$ of an electron in the symmetric double-well potential along the $x$-axis going through the potential minima. Here $\varepsilon_k$ are the energies of the stationary states $|k\rangle$; $V$ is the energy separation between the excited states 2 and 3; $\omega$ is the frequency of the laser pulse; $\delta_2$ and $\delta_3$ are the frequency detunings from the levels 2 and 3 respectively.

\vskip 4mm

Fig. 2. Probability $P_1$ of the $|0\rangle \rightarrow |1\rangle$ electron transition versus the duration $t$ of the laser pulse tuned to the resonance with the lower excited state. The values of the model parameters are (see text) $V=0.01\omega$ and $\lambda=0.001\omega$. The solid line is the approximate analytical solution (\ref{eq14}), the dots are the results of numerical calculations beyond RWA ($\lambda_{00}=10\lambda$, $\lambda_{01}=0$, $\lambda_{02} = \lambda_{03} = \lambda_{12} = - \lambda_{13} = \lambda$, $\lambda_{11} = -10 \lambda$, $\lambda_{22} = \lambda_{33} = 0$,
$\lambda_{23} = 10 \lambda$, see text).

\vskip 4mm

Fig. 3. The maximal probability of the resonant electron transition
$P_{max} = P_1(t_{op}^{res} = \pi \sqrt{2} / \lambda)$ versus
$V/\lambda$ at $V=0.01\omega$. The values of $\lambda_{km}$
are the same as in Fig. 2. The solid line is the approximate analytical solution (\ref{eq15}), the dots are the results of numerical calculations beyond RWA.

\vskip 4mm

Fig. 4. Probability $P_1$ of the $|0\rangle \rightarrow |1\rangle$ electron transition versus the duration $t$ of the nonresonant laser pulse with detunings
$\delta_2 = -\delta_3 = V/2$. The values of $V$ and $\lambda_{km}$ are the same as in Fig. 2. The solid line is the approximate analytical solution $P_1(t) = \sin^2(\lambda^2 t / V)$, the dots are the results of numerical calculations beyond RWA.

\vskip 4mm

Fig. 5. The dots show the maximum probability $P_{max}$ of the nonresonant transition $|0\rangle \rightarrow |1\rangle$ versus $V/\lambda$ at $\delta_2 = - \delta_3 = V/2$ and $V=0.01\omega$ calculated numerically beyond RWA. The values of $\lambda_{km}$ are the same as in Fig. 2. Approximate analytical solution is $P_{max} \approx 1$ at $\lambda \ll V$, see Eq. (\ref{eq16}).

\vskip 4mm

Fig. 6. Probability $P_1$ of the $|0\rangle \rightarrow |1\rangle$ electron transition versus the duration $t$ of the quasiresonant laser pulse with detunings $\delta_2 = -\delta_3 = V/2$ at $\lambda = V \sqrt{6} / 4$. The values of $V$ and $\lambda_{km}$ are the same as in Fig. 2. The solid line is the approximate expression obtained from Eqs. (\ref{eq18}) and (\ref{eq19}). The dots are the results of numerical calculations beyond RWA.

\newpage

\begin{figure}[htbp]
    \begin{centering}
        \includegraphics[width=\textwidth]{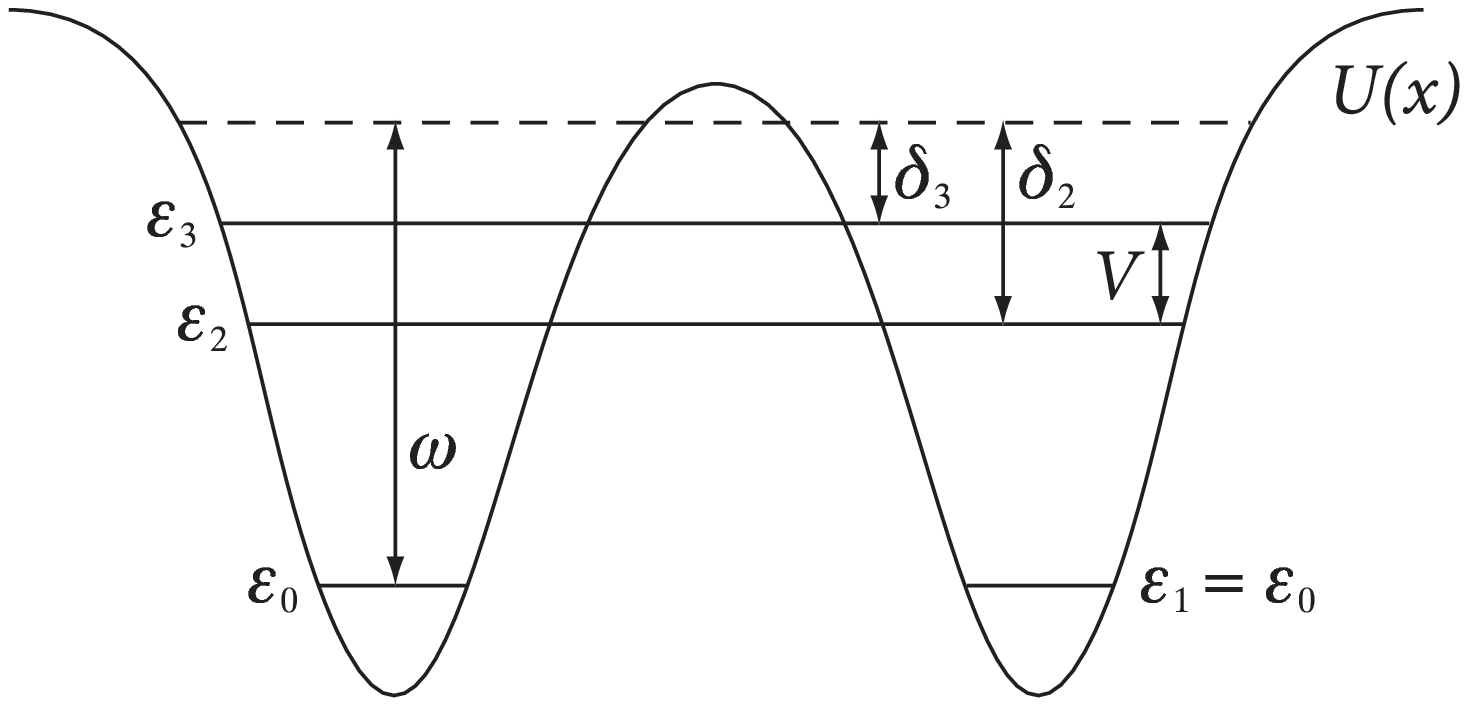}
        \label{Fig1}
    \end{centering}
\end{figure}

\vskip 40mm
\centerline{Fig.1}

\newpage

\begin{figure}[htbp]
    \begin{centering}
        \includegraphics[width=\textwidth, clip=false]{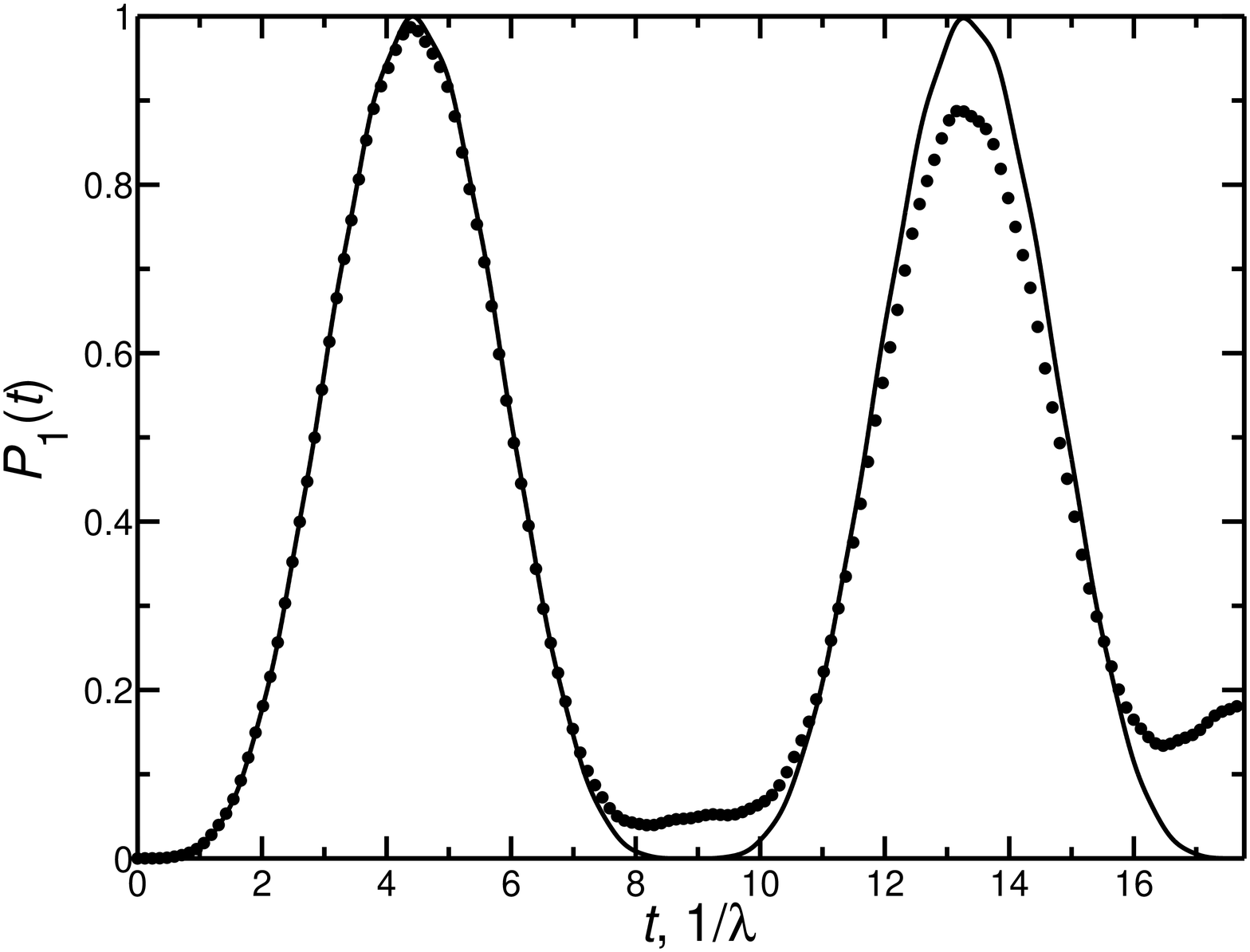}
        \label{Fig2}
    \end{centering}
\end{figure}

\vskip 40mm
\centerline{Fig.2}

\newpage

\begin{figure}[htbp]
    \begin{centering}
        \includegraphics[width=\textwidth, clip=false]{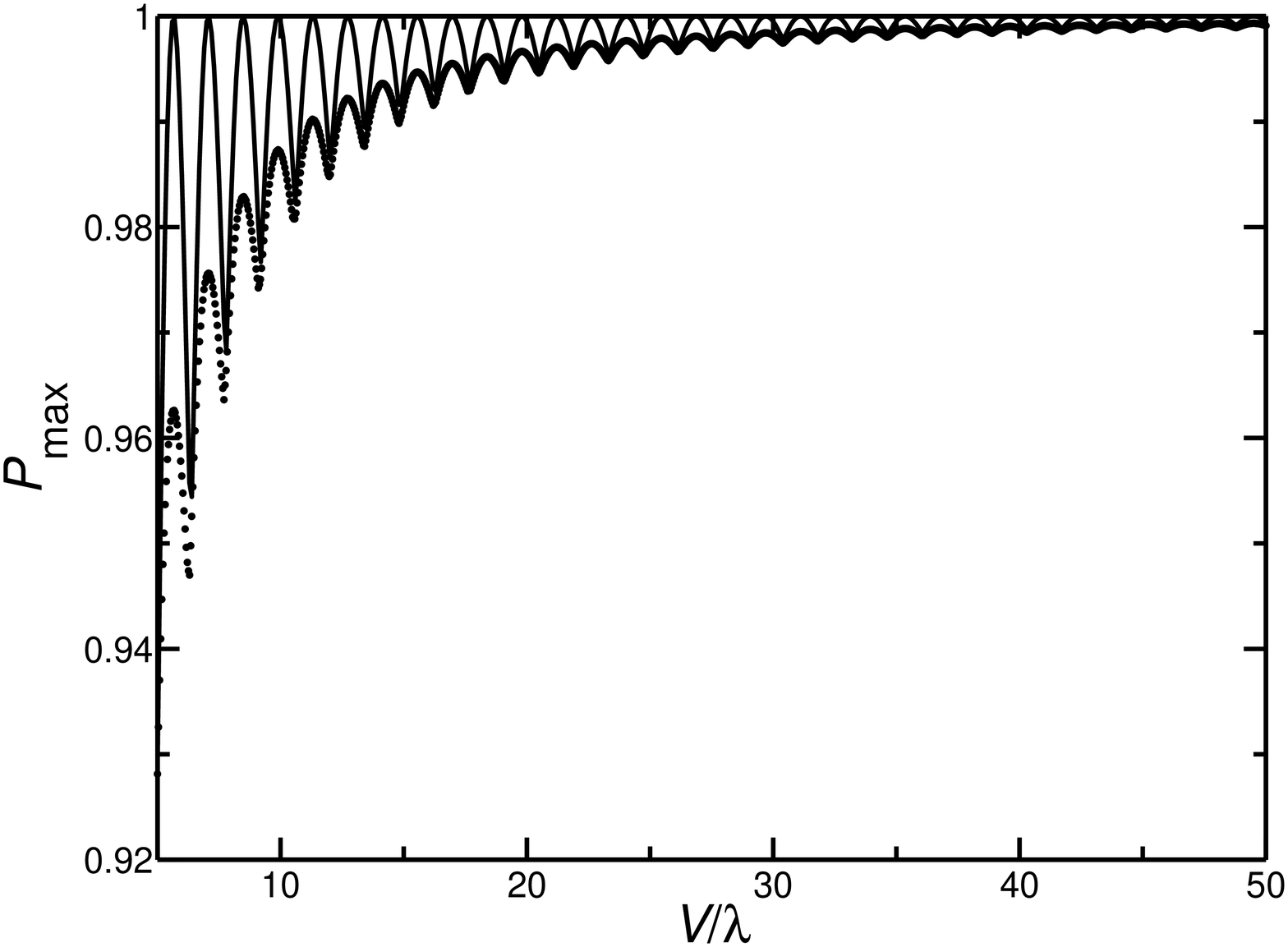}
        \label{Fig3}
    \end{centering}
\end{figure}

\vskip 40mm
\centerline{Fig.3}

\newpage

\begin{figure}[htbp]
    \begin{centering}
        \includegraphics[width=\textwidth, clip=false]{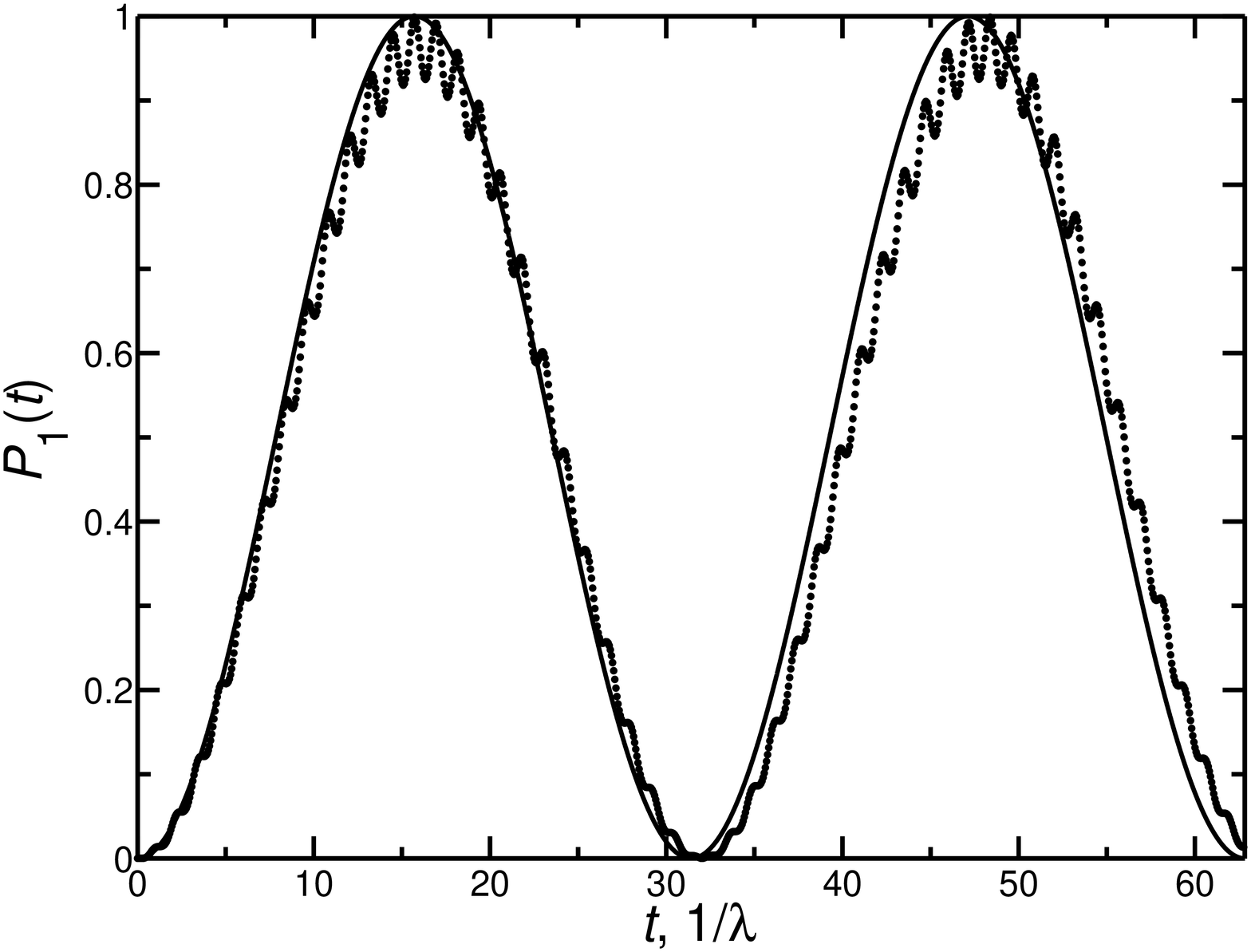}
        \label{Fig4}
    \end{centering}
\end{figure}

\vskip 40mm
\centerline{Fig.4}

\newpage

\begin{figure}[htbp]
    \begin{centering}
        \includegraphics[width=\textwidth, clip=false]{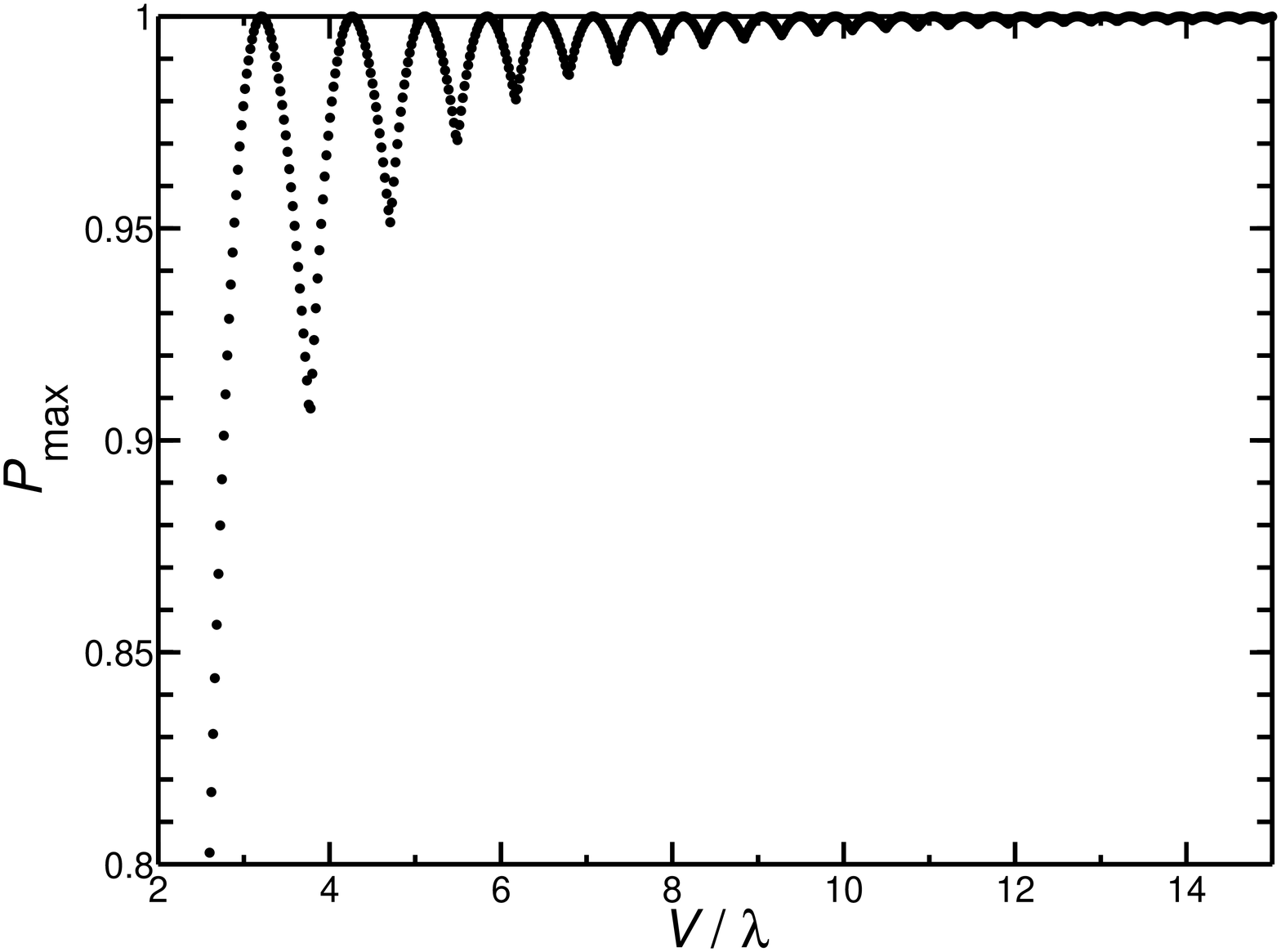}
        \label{Fig5}
    \end{centering}
\end{figure}

\vskip 40mm
\centerline{Fig.5}

\newpage

\begin{figure}[htbp]
    \begin{centering}
        \includegraphics[width=\textwidth, clip=false]{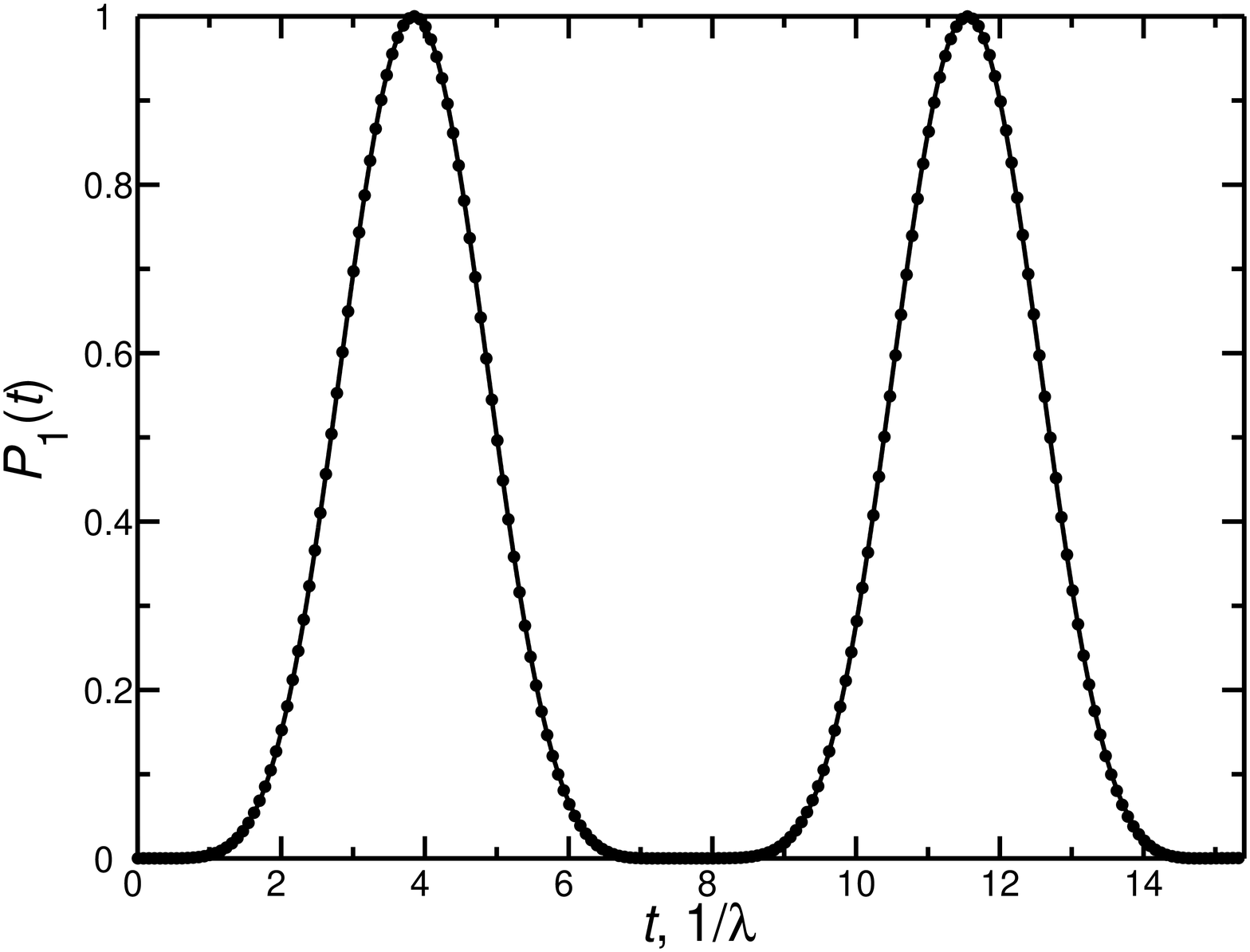}
        \label{Fig6}
    \end{centering}
\end{figure}

\vskip 40mm
\centerline{Fig.6}

\end{document}